\documentclass[journal, 10pt]{IEEEtran}

\usepackage[usenames]{color}

\usepackage{cite, amsmath,amsfonts,amsthm,amssymb}	
\usepackage{epsfig,graphicx,subfigure}	
\usepackage{extarrows}
\usepackage[colorinlistoftodos]{todonotes}

\title{Recursion-based Analysis for Information Propagation in Vehicular Ad Hoc
Networks}

\author{
  \IEEEauthorblockN{
    Minming Ni\IEEEauthorrefmark{1}, 
    Jianping Pan\IEEEauthorrefmark{2},
    Miao Hu\IEEEauthorrefmark{1}, 
    Zhangdui Zhong\IEEEauthorrefmark{1}
   } 

\IEEEauthorblockA{
 	\IEEEauthorrefmark{1}State Key Lab. of Rail Traffic Control and Safety, 
    Beijing Jiaotong University, Beijing, China}

 \IEEEauthorblockA{
 	\IEEEauthorrefmark{2}University of Victoria, Victoria, BC, Canada} 
}

\begin{document}
\pagestyle{empty} 

\maketitle
\thispagestyle{empty} 

\begin{abstract}
Effective inter-vehicle communication is fundamental to a decentralized
traffic information system based on Vehicular Ad Hoc Networks (VANETs). 
To reflect the uncertainty of the information propagation, most of the
existing work was conducted by assuming the inter-vehicle distance follows some
specific probability models, e.g., the lognormal or exponential distribution,
while reducing the analysis complexity. 
Aimed at providing more generic results, a recursive modeling framework is
proposed for VANETs in this paper when the vehicle spacing can be 
captured by a general i.i.d. distribution. 
With the framework, the analytical expressions for a series of
commonly discussed metrics are derived respectively, including the mean,
variance, probability distribution of the
propagation distance, and expectation for the number of vehicles
included in a propagation process, when the transmission failures are
mainly caused by MAC contentions.
Moreover, a discussion is also made for demonstrating the efficiency of the
recursive analysis method when the impact of channel fading is also considered. 
All the analytical results are verified by extensive simulations. 
We believe that this work is able to potentially reveal a more insightful
understanding of information propagation in VANETs by allowing to evaluate the
effect of any vehicle headway distributions. %

\end{abstract}


\section{Introduction}

Due to the potential to disseminate the safety warnings and traffic information
for significantly decreasing the number of road accidents, the Vehicular Ad Hoc
Networks (VANETs) are widely recognized as one of the few core components for
the next-generation Intelligent Transportation Systems (ITS) \cite{li}.
To further push forward VANETs' development, the US Federal
Communications Commission (FCC) has allocated 75 MHz of Dedicated Short-Range
Communications (DSRC) spectrum at 5.9 GHz to be used exclusively for
Vehicle-to-Vehicle (V2V) and Vehicle-to-Road Infrastructure (V2R) communications.
Besides, IEEE has also completed the standardization process for IEEE 1609.1,
1609.2, 1609.3, and 1609.4 for the Wireless Access of Vehicle Environments (WAVE),
which utilizes IEEE 802.11p to handle the media access control issues uniquely
happened in the VANET scenario.
Moreover, a considerable amount of VANET-oriented projects have been initiated  
by governments, automotive industry, highway management authorities, and safety
organizations, e.g., the  Vehicle Safety Consortium (VSC) in USA, the Car-to-Car
Communications Consortium (C2C-CC) sponsored by the European Union, and the
Advanced Safety Vehicle Program (ASV) in Japan.

Among all the research topics in VANETs, the performance of information
propagation in the dynamic network scenario, which is fundamental to ITS, is
always treated with a high priority.
Generally, the difficulties for the information propagation related studies come
from the time varying vehicle mobility, the burst-style data traffic loads, and
the extremely complicated radio environments.
Regarding all these issues, \cite{Yang:2003} conducted numerous simulation
studies for the information propagation distance. After that, the information
propagation issue was further studied in \cite{Jin:2006} which developed
numerical method to recursively calculate the probability of successful
propagation. With the similar idea of recursive analysis, \cite{Zhuang:JSAC11}
studied the feature of cluster size in VANETs, and proposed a
time/location-critical framework specifically for the emergency message
dissemination. In one of the most recent work \cite{Zhang:TISS14}, the
information propagation issue was investigated when the vehicles in the network
could be categorized into a number of speed distribution-determined traffic
streams. 
Although such work advanced the understanding for the information propagation
process in VANETs, most of it were confined to the condition that the vehicle
presence on a road segment follows some specific random processes, e.g., the
commonly used homogeneous Poisson point process. These carefully selected
probability models can significantly reduce the analysis complexity. 
However, it is often argued that they are in violation of the realistic vehicle spatial
distribution when some factors such as the driver behavior, traffic condition, 
and road type are considered \cite{Meng}.
%

Inspired by \cite{wang_process_2010}, this paper presents a recursive analytical
model for the information propagation process by looking into the physical
meaning of the expected propagation distance and incorporating the possible
factors for a transmission failure, when the vehicle distance headway can be
described by a general probability distribution.
Considering its generality, this study would offer more insights and enable a
more robust design of VANETs-based ITS by allowing for analytical verification
of various different headway distributions on successful information
propagation.
The remainder of this paper is organized as follows. The system model for our
recursive analysis is given in Section II. After that, the information
propagation distance's stochastic characteristics are derived in Section III. 
The verifications of our analytical results are shown in Section IV. 
Moreover, a discussion is also made in Section V to briefly describe how the
recursive model could be used when the impact of channel fading is considered.
Finally, Section VI concludes this paper.

\section{System Model}

In our analysis, the random distance $H$ between direct neighboring vehicles,
which is termed as the \emph{headway} distance, follows a general distribution whose
density function is denoted as $f_H(x)$ with expectation $\mu_H$ and variance
$\sigma_H^2$. Without considering the impact of specific routing or 
transmission schemes on information propagation in VANETs, this paper focuses on the case
that each vehicle only attempts to relay the received information via its direct
neighbor. If the trial failes after a certain retry limit or without retry due
to broadcast, the entire transmission process is
terminated, and the total propagation distance $D$ equals the distance from the
last receiving vehicle to the origin of the information. 
%
%
Due to the space limit, in this work, we mainly consider the case that the 
transmission failures are only caused by the MAC contentions. However, based on
the proposed recursive analytical framework, more general cases can also be
investigated. To demonstrate the framework's effectiveness, a discussion
section is arranged at the end of this paper,
which includes some brief results for the situation when the channel fading
becomes the major cause for the interruption of an information propagation
process. A comprehensive version of the more general analyses will be
finished as a follow-on work for this paper in the near future.


%
%
When the MAC contentions are the only concerns, it is common to 
assume that each vehicle has the identical transmission range $L$. With this
constant transmission range model, a communication can be initiated if the
distance between two vehicles is less than $L$.
However, due to the design of the backoff algorithm in MAC
schemes, it is possible that more than one vehicle may attempt to
utilize the channel at the same time, which will lead to a transmission
collision if within an interference range. This is common when the network
density or traffic load is high at some
specific time or locations. 
There is a lot of existing work focusing on the reliability of VANETs MAC,
e.g., the packet reception rate in the IEEE 802.11-based VANETs was derived in
\cite{Ma:11TVT}, and the analytical model was further improved in \cite{Ni:12GC}
for more accurately describing the \emph{Frozen Period(s)} in a channel
contention process.
For this paper, the probability for a successful
reception when the MAC contentions are the only concerns is denoted as
$p_\mathsf{s}$, and treated as a known parameter. Moreover, once the
analytical expressions for the characteristics of the information propagation are
obtained, $p_\mathsf{s}$ could be replaced by the already derived
closed-form or empirical expressions, which allow us to
reveal a more insightful understanding about the relationship between the
system parameters and the comprehensive network performance metrics.
%

\section{Theoretical Analysis}\label{sec:ana}

According to the system model, it is clear
that the information propagation is a typical renewal process.
%
In other words, if the expectation of the propagation distance $D$ can be
obtained as $\mu_D$, then each vehicle should have the same potential to
\emph{further propagate} the information forward with expectation $\mu_D$. %
Interestingly, this interpretation of the propagation distance's expectation
allows us to look into the characteristics of $D$ from a recursive aspect.
With the headway distribution, the probability for the first receiving vehicle
to be located at distance $\tau$ to a tagged transmitting vehicle is $f_H(\tau)
\,\mathrm{d}\tau$. 
For any receiving vehicle, it has probability $p_\mathsf{s}$ to successfully
finish the information reception, and 
has the potential to further relay the information forward for an expected
distance $\mu_{D}$. Hence, we could have the following recursion:
\begin{equation}
 	\mu_{D} = \int_0^{L} \left(\tau + \mu_D \right)
	f_{H}(\tau) \, p_\mathsf{s} \,\mathrm{d}\tau\quad. \label{eq:recur-expect-1}
\end{equation}
Based on (\ref{eq:recur-expect-1}), it is easy to obtain the following theorem.
\newtheorem{theorem}{Theorem}
\begin{theorem} \label{theo:1}
When the successful transmission probability determined by the MAC contention
can be presented by $p_\mathsf{s}$, the expected information propagation 
distance is given by 
	\begin{equation}
	\mu_{D} = \frac{p_\mathsf{s} \int_0^{L} \tau f_{H}(\tau)\,\mathrm{d}\tau}
	{1-p_\mathsf{s} F_{H}(L)}\quad, \label{eq:exp_a}
	\end{equation}
where $F_H(\cdot)$ is the cumulative distribution function (CDF) of the headway
distance $H$. 
\end{theorem}
The proof of Theorem 1 is quite straight-forward, which is ignored here. An
obvious but also
interesting observation of Theorem \ref{theo:1} is that, when the successful
transmission probability could be treated as independent of the transmission
distance, the expected information propagation distance is irrelevant to the
probability distribution of vehicle headway \emph{beyond} the transmission range
$L$. As will be seen later, the variance of the propagation distance under this
situation is not affected by the headway distribution beyond $L$ either.
This feature would be useful when we are trying to acquire a headway
distribution from a tremendous amount of field test data, and further using it
for calculating the stochastic characteristic of the information propagation
distance.

It is clear that there still exists an integral in the numerator of 
(\ref{eq:exp_a}), which might not be easy to be calculated when $f_H(\cdot)$ is
involved with some complicated functions. To reduce the reliance  on numerical
calculations, some attempts can be made to develop a bound estimation of the
expected propagation distance.
\newtheorem{corol}{Corollary}
\begin{corol} \label{coro:exp_bound}
The bound of the expected information propagation distance could be given as 
\setlength{\arraycolsep}{0.0em}
\begin{eqnarray}
	\frac{p_\mathsf{s} \mu_H}{1-p_\mathsf{s} F_H(L)}
	- \frac{\sqrt{\sigma_H^2 + (L-\mu_H)^2} - (L-\mu_H)}{2(1-p_\mathsf{s} F_H(L))} \notag\\
	\leq \mu_{D} \leq \frac{p_\mathsf{s}\mu_H -  p_\mathsf{s}L + p_\mathsf{s}L F_H
	(L)}{1-p_\mathsf{s} F_H(L)}\quad. \label{eq:coro-1}
\end{eqnarray}
\setlength{\arraycolsep}{5pt}
\end{corol}
\begin{IEEEproof}
The left part of the inequality is based on the results in \cite{Gallego:92}:
for any random variable $Z$ with PDF $f(z)$, mean $\mu$, and finite variance
$\sigma^2$, the following holds
\begin{equation}
	\int_z^{\infty} t f(t) \mathrm{d} t \leq \frac{\sqrt{\sigma^2 + (z-\mu)^2} + (z
	- \mu)} {2}\quad.
\end{equation}
Hence, 
\setlength{\arraycolsep}{0.0em}
\begin{eqnarray}
	\int_0^{L} \!\!\!\!\tau f_H(\tau)\mathrm{d}\tau &{}={}& 
	\int_0^{\infty} \!\!\!\!\tau f_H(\tau) \mathrm{d} \tau - \int_L^{\infty}
	\!\!\!\!\tau f_H(\tau) \mathrm{d} \tau \label{eq:exp_bound_a_1} \\
	&\geq& \mu_H - \frac{\sqrt{\sigma_H^2 + (L-\mu_H)^2} - (L-\mu_H)}{2}~.\quad
	\label{eq:coro-proof-1}
\end{eqnarray}
Besides, from (\ref{eq:exp_bound_a_1}) it is also clear that
\begin{eqnarray}
	\int_0^{L} \tau f_H(\tau)\mathrm{d}\tau 
	&{}\leq{}& \int_0^{\infty}\tau f_H(\tau)\mathrm{d}\tau 
	- L \int_L^{\infty} f_H(\tau)\mathrm{d}\tau \notag\\
	&=& \mu_H - L (1-F_H(L))\quad.\label{eq:coro-proof-2}
\end{eqnarray}
By combining the results of (\ref{eq:coro-proof-1}) and (\ref{eq:coro-proof-2}),
the bounds of $\mu_{D}$ could be obtained.
\setlength{\arraycolsep}{5pt}
\end{IEEEproof}
The advantage of having such bounds relies on the fact that,
with an arbitrary headway distribution, the expected propagation distance could
be easily estimated by the mean and variance of the vehicle headway, both of
which can be easily obtained with field data. This will be very helpful, if some
rough and quick estimations are needed. 

Meanwhile, the variance $\sigma_D^2$ of the successful propagation distance is measured as
follows:
\begin{theorem}
Given the successful transmission probability $p_\mathsf{s}$, the variance of
information propagation distance $D$ can be calculated by
	\begin{equation}
		\sigma_D^2 = \frac{p_\mathsf{s} \int_0^L \tau^2 
		f_H(\tau)\, \mathrm{d}\tau} {1 - p_\mathsf{s}F_H(L)} 
		+ \mu_D^2 \,p_\mathsf{s}\,\frac{1 - F_H(L)} {1-p_\mathsf{s}F_H(L)} \quad.
		\label{eq:theo-2}
	\end{equation}
\end{theorem}
\begin{IEEEproof}
According to the variance decomposition formula (a.k.a. the Eve's Rule), 
the recursion for $\sigma_D$ can be given as below conditional on
the distance between the first transmitter and receiver is $\tau$
\begin{equation}
\sigma_D^2 = 	\mathbf{V}(D) =  \mathbf{V}\!\left[ \mathbf{E}\!\left(D|\tau\right)
\right] + \mathbf{E}\!\left[ \mathbf{V}\!\left(D|\tau\right)
\right] \label{eq:theo-2-a}\quad.
\end{equation}
According to the mathematical definition of a random variable's
variance $\mathbf{V} (X) = \mathbf{E} \left[
(X -
\mu_X)^2\right]$, the first part in (\ref{eq:theo-2-a}) can be expanded based on
its actual physical meaning as 
\setlength{\arraycolsep}{0.0em}
\begin{eqnarray}
	\mathbf{V}\!\left[ \mathbf{E}\!\left(D|\tau\right)\right] 
	&{} = {}& \int_0^{\infty} (\mathbf{E}\left[\mathbf{E}(D|\tau) -
	\mathbf{E}(D)\right])^2
	f_H(\tau) \,p_\mathsf{s} \,\mathrm{d}\tau \notag\\
	& = & \int_0^{L} (\tau + \mu_D - \mu_D)^2 f_H(\tau) \, p_\mathsf{s}
	\,\mathrm{d}\tau \notag\\ && 
	+ \int_L^{\infty} (0 - \mu_D)^2 f_H(\tau) \, p_\mathsf{s}
	\,\mathrm{d}\tau \notag\\
	& = & p_\mathsf{s}\left(
	\int_0^L \!\!\!\!\tau^2 f_H(\tau) \,\mathrm{d} \tau + \mu_D - \mu_D^2 F_H(L)
	\right)~.~\quad~
\end{eqnarray}
Similarly, 
\setlength{\arraycolsep}{5pt}
\begin{equation}
	\mathbf{E}\!\left[ \mathbf{V}\!\left(D|\tau\right) \right] 
	= \int_0^L \!\!\!\!\mathbf{V}(D) f_H(\tau) p_\mathsf{s} \,\mathrm{d} \tau 
	= p_\mathsf{s} \mathbf{V}(D) F_H(L)\quad.
\end{equation}
Therefore, the recursion in (\ref{eq:theo-2-a}) can be simplified to
\begin{equation}
	\sigma_D^2  = p_\mathsf{s}\left(
	\int_0^L \!\!\!\!\tau^2 f_H(\tau) \,\mathrm{d} \tau + \mu_D - \mu_D^2 F_H(L)
	+ \sigma_D^2 F_H(L) \right).
\end{equation}
Finally, $\sigma_D^2$ can be directly presented as shown in (\ref{eq:theo-2}),
hence Theorem 2 is proved. 
\end{IEEEproof}

With the similar method we used for deriving $\mu_D$'s bounds, the following
corollary could be obtained.
\begin{corol}\label{coro:2}
Given $p_\mathsf{s}$ and the vehicle headway distribution, the variance of the
information propagation distance can be estimated as
	\setlength{\arraycolsep}{0.0em}
	\begin{eqnarray}
	\mu_D^2 p_\mathsf{s} \frac{1 - F_H
	(L)}{1-p_\mathsf{s}F_H(L)}  \leq  
	\sigma_D^2 &{} \leq {}& \frac{p_\mathsf{s} L \mu_H - p_\mathsf{s}L^2\left(1-F_H
	(L)\right)}
	{1 - p_\mathsf{s}F_H(L)}  \notag\\
	&& +  \mu_D^2 p_\mathsf{s} \frac{1 - F_H
	(L)}{1-p_\mathsf{s}F_H(L)} \label{eq:coro-2} \quad.
	\end{eqnarray}
\setlength{\arraycolsep}{5pt}
\end{corol}
\begin{IEEEproof}
First, the inequality shown below is easy to be obtained.
%
\setlength{\arraycolsep}{0.0em}
\begin{eqnarray} 
	\int_0^{L} \tau^2 f_H (\tau) \,\mathrm{d} \tau &{} \leq {}& 
	L \int_0^L \tau f_H(\tau) \,\mathrm{d} \tau \notag \\
	&=& L \mu_H - L \int_L^{\infty} \tau f_H(\tau) \,\mathrm{d} \tau \notag\\
	& \leq & L\mu_H - L^2(1-F_H(L)) \quad. \label{eq:coro2-zoom}
\end{eqnarray}
Therefore, by replacing the integral $\int_0^L \tau^2 f_H(\tau) \,\mathrm{d}
\tau$ in (\ref{eq:theo-2}), the right side of (\ref{eq:coro-2}) is proved. 
The left side of (\ref{eq:coro-2}) is obvious according to (\ref{eq:theo-2}).
Hence, the Corollary 2 is proved.
\setlength{\arraycolsep}{5pt}
\end{IEEEproof}

Besides the expectation and standard deviation of the information propagation distance,
the number $N$ of vehicles included in an information propagation process, which is
often termed as the \emph{Cluster Size} in the VANETs
literature,
can also be derived by the recursive analysis. The recursion for the
expectation of $N$ could be written as follows
\begin{equation}
	\mu_N  = \int_{0}^{L} \left(1 + \mu_N\right)f_H(\tau) 
	\,p_\mathsf{s} \,\mathrm{d} \tau \quad ,
\end{equation}
and another new theorem can be obtained. 
\begin{theorem}
The expected number of vehicles included in a single information propagation
process can be calculated by the successful reception probability
and the headway distribution as 
\begin{equation}
	\mu_N = \frac{p_\mathsf{s}F_H(L)}{1-p_\mathsf{s}F_H(L)}\quad.
\end{equation}
\end{theorem}

Moreover, the recursive method can also be used to derive the probability
distribution of $D$, as
\begin{theorem}
Denote function $F_D(s)$ as the CDF of the
information propagation distance $D$, it can be recursively calculated as
\setlength{\arraycolsep}{0.0em}
\begin{eqnarray}
	F_D(s) = 
	\begin{cases}
	1- p_\mathsf{s} F_H(L)~, & \!\!\!\!s = 0 \\
	\begin{aligned}
	& 1- p_\mathsf{s}F_H(L) - (1+p_\mathsf{s})F_H(s) \notag\\
	&\quad~~ + \int_0^s \!\! f_H(\tau)F_D(s-\tau)\, \mathrm{d}\tau
	\end{aligned}~, & \!\!\!\! 0 < s\leq L \\
	1- F_H(L) + \int_0^L \!\!\! f_H(\tau) F_D(s-\tau) \, \mathrm{d}\tau~,
	& \!\!\!\!s > L
	\end{cases} \label{eq:theo-distri}
\end{eqnarray}
\setlength{\arraycolsep}{5pt}\label{theo:distri}
\end{theorem}

\begin{figure}[t]
	\centering
	\includegraphics[scale = 0.8]{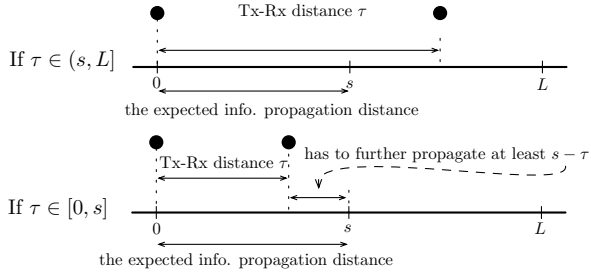}
	\caption{Calculation of information propagation distance's distribution}
	\label{fig:distri}
\end{figure}

\begin{IEEEproof} The three cases listed in the piecewise-function need to be
discussed separately. When $s = 0$, it is easy to see that
\setlength{\arraycolsep}{0.0em}
	\begin{eqnarray}
		F_{\!D}(0) &{} = {}& 
		\Pr\{ D < 0 \} = 1 - \Pr\{ \text{the 1$^{\text{st}}$ trans. is successful}
		\}\notag\\
		& = & 1 - F_H(L)p_\mathsf{s}\quad.
	\end{eqnarray}
For the case 
$0 < s \leq L$, $F_{\!D}(s) = 1- \Pr \{D > s\}$. As shown in Fig.~\ref{fig:distri}, if the
distance between the origin and the first receiver is within the range $[s,L]$,
the information propagation distance will always be longer than $s$; however, if
the first receiver is located at location $\tau \in [0,s]$, then the information
has to be further propagated at least distance $s-\tau$ to make sure $D > s$.
Therefore, 
\begin{eqnarray}
	F_{\!D}(s) &{} = {}& 1- \Pr \{D>s\} \notag\\
	& = & 1- \!\left(\!p_\mathsf{s} \left( F_H(L) - F_H(s)\right) + \!\int_0^s
	\!\!\!\!
	f_H (\tau) \overline{F_{\!D}}(s-\tau) \mathrm{d}\tau \!\right) \notag\\
	&=& 1- p_\mathsf{s}F_H(L) - (1+p_\mathsf{s})F_H(s) \notag\\
	 && + \int_0^s \!\! f_H(\tau)F_D(s-\tau)\, \mathrm{d}\tau\quad,
\end{eqnarray}
where $\overline{F_{\!D}}(\cdot)$ is the complementary cumulative distribution
function (CCDF) of the propagation distance $D$. 
Finally, when $s > L$, the first vehicle in the information propagation process
should be located within the origin's transmission range $L$, and should relay
the information with distance at least $s - L$, whose probability is
$\overline{F_{\!D}}(s-L)$, hence, 
\setlength{\arraycolsep}{5pt}
\begin{equation}
	F_{\!D}(s) = 1 - \int_0^L f_H(\tau)\,\overline{F_{\!D}}(s-L) \,\mathrm{d} \tau
	\quad.
\end{equation}
By combining all the above three parts of results, Theorem \ref{theo:distri} is
proved. 
\end{IEEEproof} 

\section{Evaluation}


To illustrate the effect generated by the MAC contentions,
the analytical results we obtained in \cite{Ni:12GC} are directly applied here to calculate the average transmission successful probability $p_\mathsf{s}$. 
In the simulation, the constant transmission range $L$ for each vehicle is set
to 100 m, and all the vehicles are randomly distributed along a road segment
according to a Poisson process with density $\lambda$, which means that the
vehicle headway should follow an exponential distribution with mean $1/\lambda$
and variance $1/\lambda^2$.
To demonstrate the impact of MAC parameters on information propagation, the
length  $Q$  of the transmission queue in each vehicle is changed from 1
to 3, and the normalized traffic loads, which is defined as $\beta L_\mathsf{P}
/ R_\mathsf{b}$\footnote{According to the definitions in \cite{Ni:12GC},
$\beta$ is the packet arrival rate at each vehicle, $L_\mathsf{P}$ is the
average packet length, and $R_\mathsf{b}$ is the bit rate.}, varies from 0 to
0.5. All the other basic parameters for the IEEE 802.11-based MAC scheme are 
listed in Table \ref{tab:1}. 
\begin{table}[!t] 
\renewcommand{\arraystretch}{1.2}
\caption{Basic MAC Parameters}
\centering \vspace{-0.1in}
\begin{tabular}{c|c||c|c} 
\hline
Bit Rate ($R_\mathsf{b}$) & 11 Mbps & Propa. Delay ($\delta$) & 2
$\mu$s \\
\hline
Slot Time ($\sigma$) & 20 $\mu$s & $DIFS$ & 50 $\mu$s \\
\hline
MAC Header & 224 bits & $W_0$ & 32 \\
\hline
PHY Header & 192 bits & $E[L_\mathsf{P}]$ & 8000 bits \\
\hline
\end{tabular} \label{tab:1}
\end{table}

\begin{figure}[t]
	\vspace{-0.2in}
	\centering
	\includegraphics[scale = 0.4]{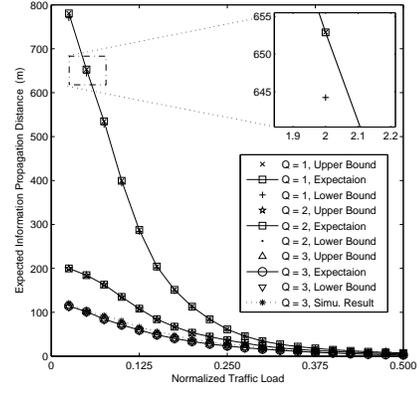}
	\vspace{-0.2in}
	\caption{Expected Information Propagation Distance vs. Normalized Traffic Load,
	while $\lambda$ = 1/5 (vehicle per meter)}
	\label{fig:mac_mean_bounds}
	\vspace{-0.16in}
\end{figure}

In Fig.~\ref{fig:mac_mean_bounds}, both the analytical and simulation results
for the expected information propagation distance are demonstrated with varying
normalized traffic loads and queuing lengths at each vehicle, while the
expectation of the headway distribution is fixed at 5 meters. 
As shown in the figure, the analytical results match the simulation results.
With the analyses in \cite{Ni:12GC}, the successful
reception rate $p_\mathsf{s}$ is expected to decrease with the increased
traffic load, which
is mainly due to the significantly increased channel contentions. Moreover, it
is also known that a longer transmission queue also reduces the average
successful reception probability. This can be explained as that a longer queue
can buffer more packets for later transmission, therefore, the probability that
a  transmission queue is empty is reduced, which indicates that network nodes
are  more likely to stay in the backoff stage rather than the idle states. 
Represented in the figure, the expected information propagation distance is
decreased with the increased traffic load and queuing length. 
Besides, the upper and lower bounds estimated from Corollary
\ref{coro:exp_bound} for different network scenarios are also illustrated in
Fig.~\ref{fig:mac_mean_bounds}. It is clear that the bounds are quite tight
comparing with the actually calculated results, which will be very useful when
the headway distribution is complicated and difficult to be handled with
numerical calculations.

\begin{figure}[t]
	\centering
	\includegraphics[scale = 0.4]{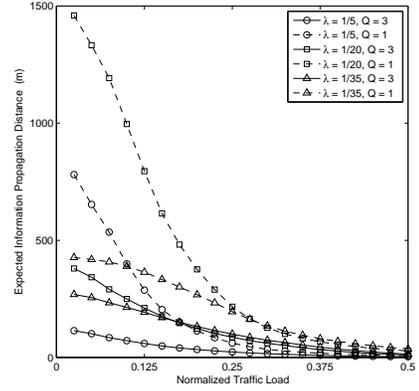}
	 \vspace{-0.2in}
	\caption{Expected Information Propagation Distance vs. Normalized Traffic Load,
	with changed $\lambda$}
	\label{fig:mac_mean_lambda}
	 \vspace{-0.2in}
\end{figure}

\begin{figure}[t]
	\centering
	\includegraphics[scale = 0.4]{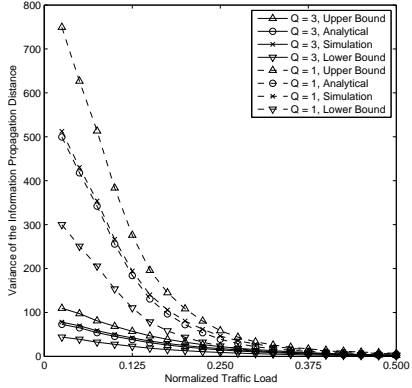}
	 \vspace{-0.2in}
	\caption{Variance of the Information Propagation Distance vs. Normalized
	Traffic Load, while $\lambda$ = 1/5 (vehicle per meter)}
	\label{fig:var}
\end{figure}

In Fig.~\ref{fig:mac_mean_lambda}, the expected information propagation distance
are illustrated with different traffic load and varied vehicular distribution
density $\lambda$. 
It is obvious that, when $\lambda$ is decreased from 1/5 to 1/20 vehicles per
meter, which results in an increase of the expected inter-vehicle distance, the
expectation of the information propagation is significantly increased. 
This is easy to be explained as that, when the hop by hop communication can be
carried out successfully with a high probability, increasing the per hop
transmission distance will directly increase the total propagation distance.
However, when $\lambda$ is further decreased, the expectation of the propagation
distance is decreased. This is due to the fact that, with a longer headway
expectation, the probability for a transmission failure is also increased, which
might start to dominate the overall performance of the information
propagation process. 
The study of the turning point of $\lambda$'s impact on the expected
propagation distance is an interesting topic, and we will dig into it in the
near future. 
It is worth to mention that, to avoid too much overlapping between different
 data sets, only the results for $Q$~=~1 and 3 are illustrated in
the figure. The changing pattern for $Q$ = 2 is similar to the ones described
above.

The analytical and simulation results for the variance $\sigma_{D}^2$ of the
information propagation distance for different network scenarios are
presented in Fig.~\ref{fig:var}, with the upper and lower bound estimations
obtained by Corollary \ref{coro:2}. 
It is clear that the analytical results and simulation results match with each
other, which validates the correctness of the newly proposed recursive analysis
method. However, comparing with the tightness between $\mu_D$ and its upper or
lower bounds demonstrated in Fig.~\ref{fig:mac_mean_bounds}, $\sigma_D$'s bound
estimations are relatively loose. 
This is mainly due to the simple mathematical relaxing techniques applied in
(\ref{eq:coro2-zoom}), which should be replaced by more sophisticated
inequalities for better tightness. This will be another follow-on work for this
paper.

\begin{figure}[t]
	\centering
	\includegraphics[scale = 0.4]{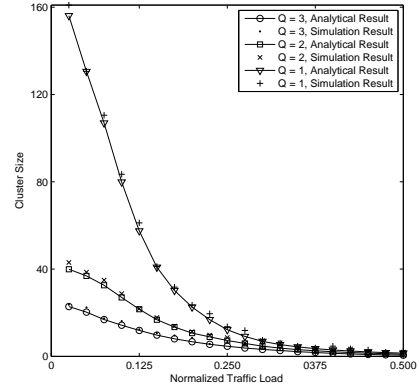}
	 \vspace{-0.2in}
	\caption{Expected Cluster Size vs. Normalized Traffic Load,
	while $\lambda$ = 1/5 (vehicle per meter)}
	\label{fig:clustersize}
	 \vspace{-0.15in}
\end{figure}

\begin{figure}[t]
	\centering
	\includegraphics[scale = 0.4]{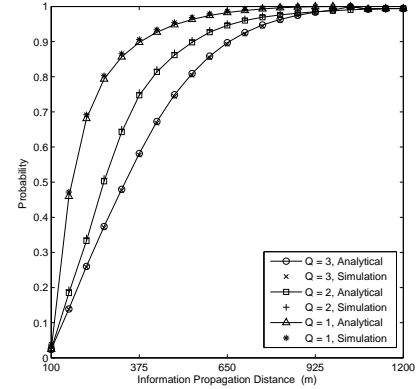}
	 \vspace{-0.2in}
	\caption{Cumulative Distribution of Information Propagation Distance 
	vs. Normalized Traffic Load,
	while $\lambda$ = 1/5 (vehicle per meter)}
	\label{fig:distribution}
\end{figure}

In Fig.~\ref{fig:clustersize} and Fig.~\ref{fig:distribution}, the expected
cluster size and the cumulative probability distribution are illustrated with
different network parameters, respectively. It is obvious that the analytical
results fit the simulation ones well. The changing pattern of the expected
cluster size is identical to the one of the expected propagation distance, which
is reasonable due to its physical meanings. For the CDF of the propagation
distance, it is clear that when the successful transmission probability
$p_\mathsf{s}$ is high, which was presented as a shorter queuing length $Q$ in
the figure, the corresponding CDF goes slower to 1, which means
$\mu_D$ will be increased. This is also coincident with the results
demonstrated in
Fig.~\ref{fig:mac_mean_bounds}.

\section{Discussion} 

When the impact of channel fading is considered, the transmission range for
each vehicle will be changed from a constant to a random variable. However, with
the recursion model described in this paper, similar analyses can also be
carried out, which will be breifly discussed in this section. 
With a specific channel model, given the
transmission distance~$\tau$, the received signal power can always be
described by a conditional probability distribution $f_{P|H}(t|\tau)$. 
%
%
%
For
example, if we assume that the channel fading follows the Rayleigh model, the
reception power's conditional PDF can be presented as 
\begin{equation}
	f_{P|\tau}(t|\tau) = \frac{1}{P_\mathsf{t} K \left( {d_0}/{\tau}
	\right)^{\alpha}}
	\exp\left(  - \frac{t}{P_\mathsf{t} K \left( {d_0}/{\tau}
	\right)^{\alpha}}\right)~,~ t>0~.
\end{equation}
where $P_\mathsf{t}$ is the transmission power, $K$ is a constant determined by
the hardware features of the transceivers, $\alpha$ is the pathloss exponent, 
and $d_0$ is the reference distance for the far-zone field.
To reduce the complexity, here the
condition for a successful data transmission is simply set to that 
the received signal power should be at least higher than the
system-determined minimum power threshold $P_\mathsf{th}$. 
However, the case when a successful
reception requires the Signal-to-Interference-Ratio (SIR) to be higher than
a threshold can also be investigated in a similar way.
With this model, the successful reception probability needs to be revised to a
transmission distance $\tau$ determined random variable as
\begin{equation}
	p_\mathsf{s}(\tau) = \int_{P_\mathsf{th}}^{\infty} f_{P|H}(t|\tau)\,
	\mathrm{d}t\quad.
\end{equation}
Then the expectation and variance of the information propagation distance could
be obtained with the following two theorems, respectively. 
\begin{theorem}
When the impact of channel fading is considered, the expected information
propagation distance $\mu_D'$ is 
\begin{equation}
	\mu_D' = \frac{\mu_H - \int_0^{\infty} \int_0^{P_\mathsf{th}}
	\tau f_{P,H}(t,\tau) \mathrm{d}t \,\mathrm{d}\tau}
	{ F_{P}(P_\mathsf{th})}\quad, \label{eq:theo-3}
\end{equation}
where $f_{P,H}(t,\tau)$ is the joint probability distribution function of the
received signal power and the transmission distance, 
and $F_P(\cdot)$ is the marginal CDF of the received signal
power.
\label{theo:3}
\end{theorem}
\begin{IEEEproof}
The recursion of the expected information propagation distance can be written as
\begin{equation}
	\mu_D' =  \int_0^{\infty} \left(\tau + \mu_D' \right)
	f_H(\tau) \int_{P_\mathsf{th}}^{\infty} f_{P|H}(t|\tau)
	\,\mathrm{d}t \, \mathrm{d} \tau\quad.
\end{equation}
With some simple manipulations, we have  
\begin{equation}
	\mu_D' = \frac{\mu_H - \int_0^{\infty} \tau f_{H}
	(\tau)\int_0^{P_\mathsf{th}} f_{P|H}(t|\tau)\,\mathrm{d}t \, \mathrm{d}\tau}
	{\int_0^{\infty} f_{H}
	(\tau)\int_0^{P_\mathsf{th}} f_{P|H}(t|\tau)\,\mathrm{d}t
	\, \mathrm{d}\tau}\quad.
\end{equation}
According to the property of the conditional probability density function, it
is clear that
\begin{equation}
	\int_0^{\infty} \!\!\!\!\tau f_{H}
	(\tau)\int_0^{P_\mathsf{th}}  \!\!\!\! f_{P|H}(t|\tau)\mathrm{d}t
	\mathrm{d}\tau
	=\int_0^{P_\mathsf{th}} \!\!\!\! f_{P}(t)\mathrm{d}t = F_{P}
	(P_\mathsf{th})~,
\end{equation}
where
\begin{equation}
	f_{P}(t) = \int_0^{\infty} f_{P|H}(t|\tau) f_{{H}}(\tau) \,
	\mathrm{d}\tau
\end{equation}
is the PDF of the received signal power when the transmission distance 
is not given, and $F_{P}(\cdot)$ is the related CDF. By combining all the above
results, (\ref{eq:theo-3}) in Theorem \ref{theo:3} is proved.
\end{IEEEproof}

\begin{theorem}
When the impact of channel fading is considered, the variance of the information
propagation distance can be calculated by 
\begin{equation}
	\sigma_D'^2 = \frac{\int_0^{\infty}\tau^2 f_H(\tau)\mathrm{d}\tau
	- \int_0^{\infty}\int_0^{P_\mathsf{th}} \tau^2 f_{P,H}(t,\tau)\,\mathrm{d}t
	\,\mathrm{d}\tau} {F_{P}(P_\mathsf{th})}~.
\end{equation}\label{theo:channel-var}
\end{theorem}
\begin{IEEEproof}
Similarly, with the Eve's Rule, and the interpretation for the physical meaning
of $\mathbf{V}\!\left[ \mathbf{E}\!\left[ D | \tau \right]\right]$ and
$\mathbf{E}\!\left[\mathbf{V}\!\left[D|\tau\right]\right]$, we could have
\setlength{\arraycolsep}{0.0em}
\begin{eqnarray}
	\mathbf{V}\!\left[\mathbf{E}\!\left[ D | \tau \right]\right] &{} = {}& 
	\int_0^{\infty} \!\!\!\!(\mu_D' + \tau - \mu_D')^2  f_H(\tau) \!\!
	\int_{P_\mathsf{th}}^{\infty} \!\!\!\!f_{P|H}(t|\tau) \,\mathrm{d}t \,
	\mathrm{d}\tau
	\notag\\
	& = & \int_{0}^{\infty} \!\!\!\!\tau^2 f_H(\tau) \, \mathrm{d} \tau 
	- \int_0^{\infty} \!\!\!\!\int_0^{P_\mathsf{th}} \!\!\!\!\tau^2 f_{P,H}(t,\tau)
	\,\mathrm{d}t \, \mathrm{d}\tau ~,\notag
\end{eqnarray}
\setlength{\arraycolsep}{5pt}
and	
\begin{equation}
	\mathbf{E}\!\left[\mathbf{V}\!\left[D|\tau\right]\right]  =   \sigma_D'^2
	\left( \int_0^{\infty}\!\!\!\! \!\!f_H(t)\,\mathrm{d}\tau - \!\!\int_0^{\infty}
	\!\!\!\!\!\!f_H(t) f_{P|H}(t|\tau) \,\mathrm{d} t \,\mathrm{d}\tau
	\right)~.
\end{equation}
By subsitituting the above two parts into (\ref{eq:theo-2-a}), Theorem
\ref{theo:channel-var} is proved.
\end{IEEEproof}

\section{Conclusions}

In this work, the stochastic characteristics of the information propagation distance
were studied with a general headway distribution by a recursion model. 
Although this paper was focused on the scenario when the MAC contentions are
the major causes for a transmission failure, a discussion was also given to
demonstrate the recursion model's adaptability when the impact of channel fading
is considered. 
As mentioned in the paper, a series of follow-on work for the recursion
model-based analysis will be conducted, including: 1) better bound
estimations for  propagation distance's variance; 2) more detailed
analysis when the SIR becomes the evaluation metric for a successful reception.
Moreover, we are also considering to further develop the recursion model for describing the unique 
store-carry-forward transmission pattern of VANETs, 
which should be another interesting topic.

\section*{Acknowledgment}
This work is partly supported by the NSFC (Grant No. 61401016), State
Key Laboratory of Rail Traffic Control and Safety (Grant No.
RCS2014ZT33, RCS2014ZQ003), and the Fundamental Research
Funds for the Central Universities (Grant No. 2014JBM154).


\end{document}